\documentclass[12pt,a4]{article}
\usepackage{amsthm,amsmath,latexsym,amssymb,amsfonts,amscd}
\usepackage{graphics,lscape,fancyhdr,array,stmaryrd,euscript}
\usepackage{ifthen,empheq}
\usepackage{sidecap}
\usepackage{ifpdf}
\usepackage{verbatim}
\usepackage{color}
\usepackage{relsize}
\numberwithin{equation}{section}
\usepackage{hyperref}
\usepackage[numbers,sort&compress]{natbib}
\usepackage[nottoc]{tocbibind}
\newcommand{\pl}{\partial}

\newcommand{\fdu}[2]{{}_{#1}{}^{#2}\,}

\begin{document}

\hfill
\vskip 0.01\textheight
\begin{center}
{\Large\bfseries 
Electromagnetic Interactions of Massive Higher-Spin Fields \\ [8pt] in 3D  via Chiral Theory}
\vskip 0.03\textheight
\renewcommand{\thefootnote}{\fnsymbol{footnote}}
Alexey \textsc{Sharapov}${}^{a}$,
David \textsc{Shcherbatov}${}^{a}$
\& Evgeny \textsc{Skvortsov}\footnote{Research Associate of the Fund for Scientific Research -- FNRS, Belgium}\footnote{Also at Lebedev Institute of Physics}${}^{b}$ 
\renewcommand{\thefootnote}{\arabic{footnote}}
\vskip 0.03\textheight
{\em ${}^{a}$Physics Faculty, Tomsk State University, \\Lenin ave. 36, Tomsk 634050, Russia}\\
\vspace*{5pt}
{\em ${}^{b}$ Service de Physique de l'Univers, Champs et Gravitation, \\ Universit\'e de Mons, 20 place du Parc, 7000 Mons, 
Belgium}\\ \vspace*{5pt}

\vskip 0.05\textheight

\begin{abstract}
We address the issue of electromagnetic interaction for massive higher-spin fields in three-dimensional Minkowski space. We show that consistent field equations can be obtained through the dimensional reduction of the higher-spin extension of self-dual Yang--Mills theory, which itself is a truncation of chiral higher-spin gravity in four dimensions. 
The resulting electromagnetic field satisfies the Bogomolny equation, and the interaction is non-minimal with the gyromagnetic ratio given by $g=1/s$, where $s$ is the spin. 
As a by-product, we obtain a new Lagrangian for free massive higher-spin fields in $3d$.

\end{abstract}

\end{center}

\section{Introduction}

Constructing interaction models for higher-spin particles is a notoriously challenging problem. For massless particles, the very possibility of such interactions is heavily constrained by the well-known Coleman--Mandula \cite{Coleman:1967ad} and Weinberg--Witten  \cite{Weinberg:1980kq} theorems.
Overcoming these no-go theorems seems impossible without compromising one or more fundamental physical principles, such as locality, unitarity, and causality. For massive higher-spin particles ($s\geq3/2$), however, the situation is somewhat more favorable. 
Massive gravity in various dimensions provides a notable example of non-trivial self-interaction for a spin-2 particle that satisfies all fundamental physical requirements \cite{de2014massive}.   In $\mathcal{N}=2$ gauged supergravity with broken supersymmetry \cite{SCHERK197960, DEWIT1985569}, a massive spin-$3/2$ gravitino can consistently propagate while being charged under the graviphoton, see also \cite{porrati2009causal, delplanque2024massive}. 
Additionally, string theory predicts an infinite tower of massive particles with arbitrarily large spin. Beyond these models and a few other results, the general landscape of massive higher-spin interactions remains largely unexplored. A number of free descriptions has been proposed, see e.g. \cite{Fierz:1939ix,Singh:1974qz,Singh:1974rc,Pashnev:1989gm,Zinoviev:2001dt}. The only general result worth mentioning here is the classification of all cubic interaction vertices in the light-cone gauge \cite{metsaev2006cubic} and, then, covariantly \cite{Metsaev:2012uy}. Some specific interactions and/or spins have also been studied \cite{Zinoviev:2006im,Zinoviev:2008ck,Zinoviev:2009hu,Zinoviev:2010cr,Buchbinder:2012xa,Buchbinder:2012iz,Zinoviev:2021cmi}.



In addition to their theoretical importance, massive higher-spin particles are of considerable phenomenological interest.
Certain atomic nuclei and hadron resonances possess high spin values. At sufficiently low energies, all these composite states can be well approximated as relativistic point particles, suggesting the existence of an effective field theory that governs their propagation in external electromagnetic and gravitational fields. Recent advances in gravitational-wave astronomy have sparked renewed interest in massive higher-spin fields.  It has been shown that the dynamics of compact binary objects, such as rotating black holes and neutron stars, can be effectively described through the scattering amplitudes of massive higher-spin fields coupled to a gravitational background 
\cite{Guevara:2018wpp,Chung:2018kqs,Guevara:2019fsj,buonanno2022snowmass, Skvortsov:2023jbn, Cangemi:2022bew,Cangemi:2022abk,Cangemi:2023bpe}.

In this paper, we primarily focus on the electromagnetic interactions of massive higher-spin fields. While the electromagnetic interaction is often regarded as the simplest and most thoroughly studied, its simplicity can be misleading when dealing with charged higher-spin fields. To date, no model has been found in $d=4$ that describes the self-consistent propagation of any massive spin in an arbitrary electromagnetic background. Models based on Fierz--Pauli Lagrangians are known to violate causality \cite{PhysRev.188.2218}, and all attempts to construct a consistent system of (non-Lagrangian) field equations fail beyond the case of a constant electromagnetic field \cite{CORTESE2014143}. Some examples of cubic electromagnetic interactions have been studied in \cite{Klishevich:1998yt,Buchbinder:2012iz}. For constant electromagnetic background, exact results have been obtained for $s=3/2$ \cite{porrati2009causal, delplanque2024massive} and $s=2$ \cite{Klishevich:1997pd}, also from string theory \cite{Argyres:1989cu,Klishevich:1998sr,Benakli:2021jxs}.

Given these limitations, a pragmatic approach is to begin with the simplest possible setup -- a strategy adopted here. First, we address the interaction problem in $d=3$, focusing on the equations of motion rather than the Lagrangian. 
Second, we extend the simplest electromagnetic background to the Bogomolny one.  
Our approach is motivated by the following observations.

In three dimensions, the D’Alembert operator can be effectively ``square-rooted'' when applied to divergence-free tensor fields, allowing free fields to be described by first-order relativistic wave equations. This approach applies to both massive and massless fields. Notably, the 3d Maxwell equations are shown to be equivalent to the Bogomolny equations, though this equivalence is established via a nonlocal field transformation. The absence of second-order derivatives in the free equations of motion significantly simplifies the subsequent analysis of consistent interactions. 
In this setup, we first define a minimal coupling of a massive spin-$s$ field to the Bogomolny background, which comprises a massless scalar field and a gauge vector field. By subsequently eliminating the scalar field using the Bogomolny equation, we can reinterpret the resulting equations of motion as describing the propagation of the massive spin-$s$ field in an electromagnetic background that satisfies the source-free Maxwell equations.  Remarkably, this non-minimal interaction is characterized by a gyromagnetic ratio of
$g=1/s$. This inverse relation between the $g$-factor and spin is consistent with the Belinfante conjecture in $d=4$.
As for nonlocalities in the background electromagnetic field, which emerge from eliminating  the scalar field, 
they can be regarded as finite-size corrections arising from the point-limit approximation of an extended charged object, such as an atomic nucleus.

The rest of the paper is organized as follows. In the next section, we provide a brief review of the interaction problem for massive higher-spin fields and outline a general method for counting the physical degrees of freedom at the level of equations of motion. In Sec. \ref{S3},  we focus on $d=3$ and introduce a consistent interaction between the massive higher-spin fields and the gauge vector field governed by the Bogomolny equation, source-free Maxwell equations, and the equations of topologically massive electrodynamics. Sec. \ref{S4} explores dual formulations of these dynamics, where duality implies that both systems of equations are dynamically equivalent and originate from a single (complex) Lagrangian. Generally, the dual system involves more fields and exhibits gauge symmetries. In Sec. \ref{S5}, we establish a connection between 3d massive theories and chiral theories in four dimensions. Specifically, we show that all the aforementioned equations and Lagrangians can be obtained through dimensional reduction from a higher-spin extension of self-dual Yang--Mills theory. 
Finally, in Sec. \ref{DC}, we summarize our findings and discuss potential avenues for future research.

\section{Review of the interaction problem} 
In the case of $4$-dimensional Minkowski space, it is customary to describe massive bosons of spin $s$ using symmetric, traceless Lorentz tensors of rank $s$: 
\begin{equation}\label{sym}
    \Phi_{\mu_1\cdots\mu_s}=\Phi_{(\mu_1\cdots\mu_s)}\,,\qquad \eta^{\mu_1\mu_2}\Phi_{\mu_1\mu_2\cdots\mu_s}=0\,.
\end{equation}
Here, the parentheses denote the total symmetrization of the enclosed indices. In the following, we adopt a condense notation
for symmetric and traceless tensors, namely $\Phi_{\mu_1\cdots\mu_s}=\Phi_{{\mu\cdots\mu}}=\Phi_{\mu(s)}$.

The free propagation of such fields is described by the Fierz--Pauli equations:
\begin{equation}\label{FP}
    T^1_{\mu(s)}:=(\square + m^2) \Phi_{\mu(s)}=0 \,, \qquad T^2_{\mu(s-1)}:=\partial^{\nu}\Phi_{\nu\mu(s-1)}=0\,.
\end{equation}
The first equation implements the mass-shell condition, while the second reduces the number of possible polarizations. Passing to the momentum representation, it is easy to see that only $2s+1$ components of the tensor $\Phi_{\mu(s)}$ are linearly independent of the others, as should be the case for a massive particle of spin $s$. Equations (\ref{sym}), (\ref{FP}) work well in all other dimensions, but for higher-dimensional spaces they do not cover all possible spins. 

From a mathematical standpoint, the Fierz–Pauli equations form an overdetermined system of linear partial differential equations, which implies the existence of nontrivial differential relations known as {\it gauge identities}:
\begin{equation}\label{GI}
    I_{\mu(s-1)}:=(\square+m^2)T^2_{\mu(s-1)}-\partial^{\nu}T^1_{\nu\mu(s-1)}=0\,.
\end{equation}
One can also see that the system of Eqs. (\ref{FP}) is {\it involutive}\footnote{In loose terms, a system is involutive if it contains no hidden integrability conditions of lower order. Any system of differential equations can be brought into an involutive form by augmenting  it with
differential consequences.  For further details, see \cite{Kaparulin:2012px, seiler_involution_2010}.}. This property enables the use of a general formula for covariant counting of physical degrees of freedom, as introduced in \cite{Kaparulin:2012px}. In the simplest case of irreducible gauge symmetries and identities, the formula reads
\begin{equation}\label{N}
    \mathcal{N}= \sum_{k=1}^\infty k(t_k-l_k-r_k)\,,
\end{equation}
where $t_k$, $l_k$ and $r_k$ represent, respectively, the number of equations, gauge identities, and gauge symmetries of order $k$. 
The order $k$ is defined as the highest order of derivatives present in a given equation, identity, or gauge symmetry transformation. For gauge transformations,  $k$ corresponds to the highest derivative acting on the gauge parameter, while for gauge identities it is given by the sum of the highest derivative of the gauge identity generator and the differential order of the equation it acts upon. Let us apply formula (\ref{N}) to the case at hand. Since there is no gauge symmetry, we have $r_k=0$. Counting the components of symmetric traceless tensors in four dimensions, we find
\begin{equation}
    t_1=s^2\,,\qquad t_2=(s+1)^2 \,, \qquad l_3=s^2\,,
\end{equation}
with the other numbers being zero. Substituting into (\ref{N}) yields:
\begin{equation}
    \mathcal{N}=1\cdot s^2+2\cdot (s+1)^2-3\cdot s^2=2(2s+1)\,.
\end{equation}
This represents the correct number of physical degrees of freedom in phase space (twice the number of physical polarizations). In three dimensions, the calculation gives $\mathcal{N}=4$, which is also the correct result, as all massive spinning particles in three-dimensional spacetime possess two polarizations, regardless of spin.

The free field equations (\ref{FP}) and their gauge identities (\ref{GI}) serve as the starting point for introducing interaction. In most cases, this can be done perturbatively with respect to a  coupling constant. The primary requirement (in addition to other physical constraints) is that the interaction should not alter the number of physical degrees of freedom. A general procedure for introducing perturbative interactions was developed in \cite{Kaparulin:2012px}. Adapted to our context, it proceeds as follows. Let $T_a(\phi)=0$ be an overdetermined system of involutive PDEs without gauge invariance and 
let $L_A^a$ denote a basis of the gauge identity generators, so that $I_A:=L_A^a T_a=0$. By definition, $(L_{A}^a)$ is a matrix differential operator. The interaction is then introduced in the form
\begin{equation}\label{DEM}
    \tilde T_a=T_a+gT_a^{(1)}+g^2 T^{(2)}_a+\ldots\,,
\end{equation}
where $g$ is a coupling constant and $T^{(k)}_a$ represent interaction vertices. The deformation of the free equations $T_a$ is accompanied by a deformation of the gauge identity generators:
\begin{equation}
      \tilde L^a_A=L_A^a+g L^{(1)a}_A+g^2 L_A^{(2)a}+\ldots \,. 
      \end{equation}
The interaction vertices $T^{(k)}_a$  and the operators $L^{(k)a}_A$ are determined by requiring that the deformed gauge identities hold, specifically:
\begin{equation}\label{DGI}
    \tilde I_A:=\tilde L^a_A \tilde T_a=0\,.
\end{equation}
For the interaction to preserve the number of physical degrees of freedom, the numbers $\tilde t_k$ and $\tilde l_k$ associated with the deformed equations of motion (\ref{DEM}) and the gauge identities (\ref{DGI}) must satisfy the following condition:
\begin{equation}\label{fc}
    \sum_{k=0}^\infty k(\tilde t_k-\tilde l_k)=\sum_{k=0}^\infty k(t_k-l_k)\,.
\end{equation}
Note that the interaction may increase the differential orders of both the field equations and the gauge identities. For  formal consistency, only the sum in Eq. (\ref{fc}) must remain invariant. 

The method outlined above for introducing interactions, though conceptually straightforward, can result in cumbersome calculations in practice. 
The first-order deformation of the Fierz--Pauli equations (\ref{FP}) by the external electromagnetic and gravitational fields was considered in 
\cite{CORTESE2014143}. In that paper, the interaction vertices were assumed to be local for background and dynamical fields. The general conclusion is that for $s>1$ such 
an interaction may exist only if the background fields are subject to severe differential constraints. 
The following table provides a rough classification of known electromagnetic interactions based on the properties of the equations of motion and the constraints imposed on the background electromagnetic field. 
The left column describes the ideal scenario, while the right column presents possible compromises.
\begin{center}
\begin{tabular}{ |p{8mm}|p{8cm}|p{8cm}| } 
\hline
Type & \hspace{3.5cm} I & \hspace{3.5cm} \bf II \\
\hline
A  & Electromagnetic background is abitrary. &   Electromagnetic fields are subject to some differential equations (e.g. Maxwell's equations and/or symmetry conditions).   \\ [1.5cm]
B&   Equations are local in the electromagnetic field.  &   Equations may contain nonlocalities in the background electromagnetic field.  \\ [0.8cm]
C& The principal symbol remains unchanged. & The principal symbol changes. In particular, equations may involve higher derivatives.\\[0.7cm]
D&     Equations are Lagrangian.    &    Equations are non-Lagrangian as they stand.   \\ 
\hline
\end{tabular}
\end{center}

We conclude this section with a few examples illustrating this classification.

\paragraph{Lower spins ($s\leq 1$).} The cases $s=0$ and $s=1/2$ correspond 
to scalar and spinor electrodynamics and pose no difficulties. Both fall entirely within the first column. Charged massive particles of spin one are usually described by the complex vector field $\Phi_\mu$ whose dynamics are governed by the Proca Lagrangian. By gauging the  $U(1)$ global symmetry, we arrive at the Lagrangian 
\begin{equation}\label{proca}
    L=-\frac14 \bar G_{\mu\nu}G^{\mu\nu}+\frac12 m^2 \bar \Phi_\mu\Phi^\mu +iegF_{\mu\nu}\bar \Phi^\mu\Phi^\nu\,,
\end{equation}
where
\begin{equation}
    G_{\mu\nu}=D_\mu \Phi_\nu-D_\nu\Phi_\mu\,,\qquad D_\mu=\partial_\mu+ieA_\mu
\end{equation}
and $A_\mu$ is a given external electromagnetic potential with strength $F_{\mu\nu}=\partial_\mu A_\nu-\partial_\nu A_\mu$.  Here, we have also added a non-minimal term proportional to $g$, which describes the magnetic moment interaction. The Lagrangian equations
\begin{equation}\label{proca1}
    D^2\Phi_\mu + m^2\Phi_{\mu} + iegF_{\mu\nu}\Phi^{\nu} - D_{\mu}D^{\nu}\Phi_{\nu} = 0
\end{equation}
are not involutive. Taking the covariant divergence, we obtain the first-order differential constraint 
\begin{equation}\label{div}
    m^2D^\mu \Phi_\mu + ie(g - 2)F_{\mu\nu}D^{\mu}\Phi^{\nu} + ie(g - 1)(\partial^{\mu}F_{\mu\nu})\Phi^{\nu} =0\,,
\end{equation}
which allows one to bring Eq. (\ref{proca1}) into the form 
\begin{equation}\label{proca2}
\begin{array}{c}
    \Big(D^2\delta_{\mu}^\nu + \displaystyle \frac{ie}{m^2}(g-2)F_{\lambda}{}^{\nu}D_{\mu}D^{\lambda}\Big)\Phi_{\nu} \\[5mm]
\displaystyle +m^2\Phi_{\mu} + iegF_{\mu\nu}\Phi^{\nu}+\frac{ie}{m^2}(g-2)(\partial_{\mu}F_{\nu\lambda})D^{\nu}\Phi^{\lambda} + \frac{ie}{m^2}(g-1)D_{\mu}\big((\partial^{\lambda}F_{\lambda\nu})\Phi^{\nu}\big) = 0\,.
\end{array}
\end{equation}
Unlike the original field equations (\ref{proca1}), the principal symbol of equation (\ref{proca2}) is non-degenerate but begins to depend on the external electromagnetic field unless $g=2$. A detailed analysis in \cite{PhysRev.188.2218} demonstrates that this interaction does not affect the causal structure of the massive field $\Phi_\mu$; for all values of $g$, the ray cone coincides with the light cone. Consequently, the Lagrangian (\ref{proca}) describes an electromagnetic interaction of type AIBICIIDI. 

\paragraph{Homogeneous electromagnetic backgrounds.} Suppose  the strength tensor $F_{\mu\nu}$ of the electromagnetic field is constant, $\partial_\lambda F_{\mu\nu}=0$. Then, for any integer spin $s$,  we can construct an interaction of type AIIBICIDII. The corresponding non-Lagrangian field equations are given by
\begin{equation}
    \label{spins}
         T^1_{\mu(s)} := (\Delta_g + m^2)\Phi_{\mu(s)} = 0\,,\qquad T^2_{\mu(s-1)} := D^{\nu}\Phi_{\nu\mu(s-1)} = 0\,,
\end{equation}
where we use the wave operator 
\begin{equation}\label{dg}
        \Delta_g\Phi_{\mu(s)} = D^2\Phi_{\mu(s)} - iegF_{\lambda \mu}\Phi^{\lambda}{}_{\mu(s-1)}\,.     
\end{equation}
introduced in \cite{deser2002inconsistencies}. 
The constant $g$ represents the gyromagnetic ratio. 
Using the identity 
\begin{equation}
\begin{array}{rcl}
[\Delta_g, D^{\nu}]\Phi_{\nu\mu(s-1)} &=& -ie(g-2)F_{\rho\lambda}D^{\rho}\Phi^{\lambda}{}_{\mu(s-1)} 
\\[5mm]
&-& ie(g - 1)(\partial^{\rho}F_{\rho\lambda})\Phi^{\lambda}{}_{\mu(s-1)} + ieg(\partial_{\rho}F_{\lambda \mu})\Phi^{\lambda\rho}{}_{\mu(s-2)}\,,
\end{array}
\end{equation}
which holds for any electromagnetic field,  we obtain
\begin{equation}
 D^{\nu}T^1_{\nu\mu(s-1)} - (\Delta_g + m^2)T^2_{\mu(s-1)} = ie(g-2)F^{\nu\lambda}D_{\lambda}\Phi_{\nu\mu(s-1)}\,.
\end{equation}
For $g=2$, we get the desired gauge identity of order three, which provides the right number of propagating modes.
The value $g=2$ is also predicted by string theory \cite{ARGYRES198989, PhysRevD.46.3529, porrati2011string, benakli2021higher} and is consistent with the gyromagnetic ratio of $W^\pm$ bosons in the Standard Model \cite{RevModPhys.36.632, holstein2006large}.  Interestingly, a charged rotating black hole is also characterized by the gyromagnetic ratio of two
\cite{ 10.1063/1.1664769, aliev2007gyromagnetic}. 

\section{Charged massive fields in three dimensions}\label{S3}
Now we specify our considerations to three-dimensional Minkowski space. Equation (\ref{FP}) still defines the propagation of free massive fields of spin $s>0$ and the electromagnetic field is described by the vector potential $A_\mu$ with the strength tensor $F_{\mu\nu}$. 

In order to introduce a non-minimal electromagnetic coupling we will apply a kind of Dirac's trick. To this end, we introduce the first-order differential operator $W$ whose action on covariant tensors is defined by the formula
\begin{equation}\label{W}
    W\Phi_{\mu_1\cdots \mu_s} = \sum_{k=1}^s\varepsilon_{\mu_k}{}^{\lambda\nu}\partial_{\lambda} \Phi_{\mu_1\cdots \nu\cdots \mu_s}\,,
\end{equation}
$\varepsilon_{\mu\nu\lambda}$ being the Levi-Civita symbol. All indices are raised and lowered with the help of the Minkowski metric $(+--)$.
The operator $W$ is formally Hermitian and has the meaning of the Pauli--Lubanski pseudoscalar. It commutes with the action of the Killing vectors and must be constant in any irreducible representation of the proper Poincar\'e group. Suppose now that $\Phi_{\mu_1\cdots \mu_s}$ is a fully symmetric and traceless tensor obeying the tranversality condition 
\begin{equation}\label{dF}
    \partial^{\mu_1}\Phi_{\mu_1\cdots \mu_s}=0\,.
\end{equation}
Then, when applied to such tensors, the operator $W$ satisfies the identity $W^2=-s^2\partial_\mu\partial^\mu $. In other words, the matrix differential operator $W$ defines a square root of the 3d d'Alembertian $\square=\partial_\mu\partial^\mu$ restricted to symmetric traceless fields obeying the transversality condition (\ref{dF}).  
In group-theoretical terms, this means that spin irreducibility implies mass irreducibility for this class of fields. Now, the free propagation of a  spin-$s$ and mass-$m$ field is described by the relativistic wave equation 
\begin{equation}\label{Wm}
    W \Phi_{\mu(s)} =\pm ms \Phi_{\mu(s)}
    \end{equation}
combined with the transversality condition\footnote{For $s=1$, the transversality condition (\ref{dF}) becomes redundant and follows from Eq. (\ref{Wm}). The latter describes the massive Chern--Simons field \cite{TOWNSEND198438}. } (\ref{dF}). Geometrically, the eigenvalue $\pm s$ of the operator $m^{-1} W$ represents the projection of the particle's spin onto the direction of its momentum.  This quantity is commonly referred to as helicity. 
Depending on the sign, we speak of right-handed and left-handed helicities.  

Note that the divergence of (\ref{Wm}) vanishes identically on the transversal fields, leading to a gauge identity of differential order $2$. Using formula (\ref{N}), we can easily find that the overdetermined system (\ref{dF}), (\ref{Wm}) describes $2$ physical 
degrees  of freedom, or equivalently, one physical polarization. This result should be compared with the $4$ physical degrees of freedom (two polarizations) for the 3d Fierz--Pauli equations (\ref{FP}).  The discrepancy in numbers is easily explained. Equation (\ref{Wm}), involving the pseudoscalar operator $W$,  violates invariance under spacetime reflections, which exchanges the two polarizations in the case of Fierz and Pauli.  More formally, each factor in the Klein--Gordon operator,
\begin{equation}
    (\square+m^2)s^2=-(W-sm)(W+sm)\,,
\end{equation}
describes the propagation of a single helicity -- either the right-handed  or left-handed.  Taking the square root requires choosing between the two possible helicities which are related by reflection. 
Thus, Eqs. (\ref{dF}),  (\ref{Wm}) define the chiral representations of the 3d Poincar'e group. 

A similar trick can also be applied to massless particles. The chiral counterpart of the Maxwell--d'Alembert equations in three dimensions is known as the Bogomolny equations. In our notation, they can be written as 
\begin{equation}\label{BE}
    WA_\mu=\partial_\mu \varphi\,.
\end{equation}
Here, $A_\mu$ is the vector potential and $\varphi$ is a scalar field. These equations are invariant under standard gauge transformations $A_\mu\rightarrow A_\mu+\partial_\mu\lambda$ and can be rewritten in terms of the strength tensor $F_{\mu\nu}$. Again, applying formula (\ref{N}) for the covariant count of degrees of freedom yields
\begin{equation}
    \mathcal{N}=1\cdot 3-1\cdot 1=2\,,
\end{equation}
where the first summand represents the three first-order equations of motion and the second corresponds to one gauge symmetry of the first order. Taking the divergence of Eq. (\ref{BE}) or applying the operator $W$ to it, we obtain the uncoupled system of the Maxwell and d'Alembert equations
\begin{equation}\label{MdA}
    \partial^\mu F_{\mu\nu}=0\,,\qquad \square \varphi=0\,.
\end{equation}
Together, these second-order equations describe the propagation of $4$ physical degrees of freedom. This allows us to interpret the Bogomolny equations (\ref{BE})
as a square root of the Maxwell--d'Alembert system (\ref{MdA}). Note that the Bogomolny equations are non-Lagrangian,  as the number of fields exceeds the number of equations.  

We now suppose that the massive field $\Phi_{\mu(s)}$ is complex and, therefore, may carry an electric charge. A minimal electromagnetic coupling of the chiral system  (\ref{dF}), (\ref{Wm}) to the Bogomolny background (\ref{BE}) is described by the following set of equations:
\begin{equation}\label{Int}
     W_A \Phi_{\mu(s)} =s(m +ie\varphi)\Phi_{\mu(s)}\,,\qquad D^{\nu}\Phi_{\nu\mu(s-1)}=0\,.
\end{equation}
Here, we introduce the $U(1)$-covariant operator $W_A$ by replacing the partial derivative in (\ref{W}) with the covariant derivative $D_\mu=\partial_\mu+ ieA_\mu$.
Note that the scalar $\varphi$ appears on the right-hand side of the first equation. It is necessary to verify that the above equations still describe the propagation of a single physical mode.  First, we observe that the field equations (\ref{Int}) have the same differential order as their free counterparts (\ref{dF}) and (\ref{Wm}).  Therefore, the only critical point is to preserve both the number and the order of the gauge identities. By writing the first equation in (\ref{Int}) as $\tilde T^1_{\mu(s)}=0$ and the second as $\tilde T^2_{\mu(s-1)}=0$, we find 
\begin{equation}\label{tI}
   \tilde I_{\mu(s-1)}:=D^{\nu} \tilde T^1_{\nu\mu(s-1)} - \big(W_A-s(m+ie\varphi)\big)\tilde T^2_{\mu(s-1)} = ies(WA^{\nu}-\partial^{\nu}\varphi)\Phi_{\nu\mu(s-1)}\,.
   \end{equation}
Thus, whenever the background fields $A_\mu$ and $\varphi$ satisfy the Bogomolny equations (\ref{BE}), system (\ref{Int}) possesses $2s-1$ gauge identities $\tilde I_{\mu(s-1)}=0$ of order two. Applying formula (\ref{N}) yields two physical degrees of freedom per point.

There is an alternative interpretation of the field equations (\ref{Int}), in which $A_\mu$ is considered a genuine electromagnetic field. 
Let $A_\mu$ be a smooth solution to the source-free Maxwell equations $\partial^\mu F_{\mu\nu}=0$. Then, by the Poincar\'e Lemma, there exists a $3$-form $\Lambda_{\mu\nu\lambda}$ such that 
\begin{equation}
    F_{\mu\nu}=\partial^\lambda\Lambda_{\mu\nu\lambda} \,.
\end{equation}
By writing $\Lambda_{\mu\nu\lambda}=\varepsilon_{\mu\nu\lambda}\varphi$ for some scalar $\varphi$ and substituting this back, we obtain the Bogomolny equation, which now becomes an identity. Given an on-shell electromagnetic field $A_\mu$, the Bogomolny equations determine $\varphi$ up to an additive constant:
\begin{equation}
    \varphi[A]=\int_0^1d tx^\mu WA_\mu (tx) +\varphi_0\,.
\end{equation}
With $A_\mu$ and $\varphi=\varphi[A]$,  Eq. (\ref{Int}) defines an electromagnetic interaction that belongs to the class AIIBIICIDII. 

In the context of gravitational interaction, the idea of introducing additional background fields that are related to the original ones via differential equations was proposed in \cite{cortese2017general}. It was demonstrated that this approach allows for a significant extension of the set of geometries that enable consistent propagation of massive higher-spin fields.  In our case, the role of such an additional background field is played by the scalar $\varphi$. 

One might wonder about the gyromagnetic ratio of a charged, spinning particle whose dynamics are governed by Eq. (\ref{Int}).
To answer this question, we apply the operator $W_A$ to both sides of the first equation. This results in the second-order differential equation
\begin{equation}
   D^2\Phi_{\mu(s)}+ M^2\Phi_{\mu(s)} -\frac{ie}{s} F_{\lambda \mu}\Phi^\lambda{}_{\mu(s-1)}=0 \,. 
\end{equation}
Looking at the first term, we conclude that the ray cone coincides with the light cone.  Therefore, the equation defines the causal propagation of the matter fields. The second term resembles the mass term, but the ``effective mass'' $M=m+ie\varphi$ varies with the point and is given by the sum of the ``bare mass'' $m$ and the ``electromagnetic mass'' $ie\varphi[A]$. Finally, the coefficient in the third term determines the gyromagnetic ratio $g=1/s$ (cf. Eq. (\ref{dg})). In four dimensions, this relationship between the $g$-factor and the spin of a charged particle was conjectured by Belinfante \cite{PhysRev.92.997}, \cite{PhysRevLett.24.1381}.


We conclude this section with a brief discussion of interaction with the topologically massive electromagnetic field \cite{SCHONFELD1981157, PhysRevLett.48.975}. The mass of the gauge vector field is generated by augmenting the Lagrangian of 3d Maxwell's electrodynamics with the Chern--Simons term\footnote{This Lagrangian belongs to an infinite family of higher-derivative extensions of Chern--Simons theory; see, for example, \cite{deser1999higher, kaparulin2015higher,abakumova2019stable}.}. The corresponding equations of motion are
\begin{equation}\label{MCS}
    (W- m)WA_\mu=0\,.
\end{equation}
Setting $m=0$, we recover the usual Maxwell theory in $d=3$. For any  mass value, the equations retain the standard gauge invariance and describe one propagating polarization of $A_\mu$.  Let us introduce the ``extended'' covariant derivative 
\begin{equation}
    \bar D_\mu=\partial_\mu +ieA_\mu + \frac{i e}{m}(W A)_\mu\,.
\end{equation}
It is easy to see that Eq. (\ref{MCS}) is equivalent to the commutativity condition $[\bar{D}_\mu, \bar{D}_\nu]=0$. By replacing the partial derivative $\partial_\mu$ in (\ref{W}) with $\bar D_\mu$, we define the $U(1)$-covariant operator $\bar W_A$, which acts in the space of complex tensor fields $\Phi_{\mu(s)}$. Coupling to the gauge vector field is now introduced through the field equations
\begin{equation}
    \bar W_A\Phi_{\mu(s)}=\pm \Phi_{\mu(s)}\,,\qquad \bar D^{\nu}\Phi_{\nu\mu(s-1)}=0\,.
\end{equation}
The commutativity of extended covariant derivatives ensures that the system describes the propagation of a single physical polarization, provided that the background field $A_\mu$ satisfies Eq. (\ref{MCS}). In the spinor formalism, these equations were first considered in \cite{doi:10.1142/S0217751X97002292}. They fall into class AIIBICIDII. 

\section{Dual equations and Lagrangians}\label{S4}
The system of equations (\ref{BE}) and (\ref{Int}) has two notable flaws. First, it violates parity, describing only half of the polarizations of a spinning particle in three dimensions. Second, the field equations are non-Lagrangian. In particular, there is no back-reaction of the matter fields on the dynamics of gauge fields. To address these shortcomings, we begin by doubling the spectrum of matter fields, considering the fields $\Phi^\pm$ with left- and right-handed helicities. 
They  are assumed to carry opposite electric charges $\pm e$, so the equations of motion take the form
\begin{equation}\label{Int2}
     W_A \Phi^{\pm}_{\mu(s)} =\pm s(m +ie\varphi)\Phi^\pm_{\mu(s)}\,,\qquad D^{\nu}\Phi^\pm_{\nu\mu(s-1)}=0\,.
\end{equation}
To make these equations Lagrangian, we need to further extend the spectrum of gauge and matter fields. Specifically, new fields $\Psi_\pm$, $\Lambda_\pm$, and $B$ are introduced as Lagrange multipliers for the field equation (\ref{BE}) and (\ref{Int2}). The Lagrangian for the extended system of fields is now expressed as
\begin{equation}\label{L}
\begin{array}{rcl}
    L_1&=&\Psi_+^{\mu(s)}\big((W_A-sm)\Phi^+_{\mu(s)}-ise\varphi\Phi^+_{\mu(s)}\big)+\Lambda_+^{\mu(s-1)}D^{\nu}\Phi^+_{\nu\mu(s-1)} \\[5mm]
&+&\Psi_-^{\mu(s)}\big((W_A+sm)\Phi^-_{\mu(s)}+ise\varphi\Phi^-_{\mu(s)}\big)+\Lambda_-^{\mu(s-1)}D^{\nu}\Phi^-_{\nu\mu(s)} \\[5mm]
   &+& B^\mu(WA_\mu-\partial_\mu\varphi)\,.
\end{array}
\end{equation}
 Similar to $\Phi^\pm_{\mu(s)}$, the new tensor fields $\Psi_\pm^{\mu(s)}$ and $\Lambda_\pm^{\mu(s-1)}$ are symmetric, traceless, and complex. Consequently, the Lagrangian is also complex. Nevertheless, it leads to meaningful equations of motion, as we will see shortly. 
Varying this Lagrangian with respect to the fields $\Psi_\pm$, $\Lambda_\pm$, and $B$ reproduces the original Eqs. (\ref{BE}, \ref{Int2}). Varying $\Phi^\pm_{\mu(s)}$, $A_\mu$, and $\varphi$, we obtain the new field equations:
\begin{equation}\label{DS}
 (W_A\mp sm)\Psi^{\mu(s)}_\pm\mp ise\varphi \Psi^{\mu(s)}_\pm - D^{\mu} \Lambda^{\mu(s-1)}_\pm = 0\,,\qquad  WB_\mu=ie(J^+_\mu - J^-_\mu)\,, \qquad \partial_\mu B^\mu=iesQ\,,     
\end{equation}
where 
\begin{equation}
    J^{\pm}_\mu: = s\varepsilon_{\mu\rho}{}^{\lambda}\Psi^{\rho\nu(s-1)}_{\pm}\Phi^{\pm}_{\lambda\nu(s-1)}  -\Lambda^{\nu(s-1)}_{\pm}\Phi^{\pm}_{\mu\nu(s-1)}\,,\qquad Q:=\Psi^{\mu(s)}_+\Phi^+_{\mu(s)} - \Psi^{\mu(s)}_-\Phi^-_{\mu(s)}.
\end{equation}
We will refer to Eqs. (\ref{DS}) as a {\it dual system}.  It is a general fact that the system dual to an involutive system of linear PDEs describes the same number of physical degrees of freedom. In the case under consideration, this can be verified directly using formula (\ref{N}). Indeed, after the complexification of the matter field $\Phi$, the combined system of Bogomolny (\ref{BE}) and spin-$s$ equations (\ref{Int}) is characterized by $2+4$ physical degrees of freedom (or $1+2$ real polarizations). The gauge identities (\ref{tI}) imply the invariance of the Lagrangin (\ref{L}) under the gauge transfromations
\begin{equation}\label{R1}
    \delta_\xi\Psi^{\mu(s)}_\pm = D^{\mu}\xi_\pm^{\mu(s-1)}\,,\qquad \delta_\xi \Lambda^{\mu(s-1)}_\pm=(W_A\mp ise\varphi\mp sm)\xi_\pm^{\mu(s-1)}\,,\qquad 
    \delta_\xi B_\nu= ies\xi_\pm^{\mu(s-1)} \Phi^\pm_{\mu(s-1)\nu}\,. 
\end{equation}
This yields $4(2s-1)$  gauge transformations of order one. By adding the $U(1)$ gauge transformations 
\begin{equation}\label{R2}
    \delta_\varepsilon A_\mu=\partial_\mu \varepsilon \,,\qquad \delta_\varepsilon \Phi^\pm_{\mu(s)} =\mp ie\varepsilon \Phi^\pm_{\mu(s)} \,,\qquad \delta_\varepsilon \Psi_\pm^{\mu(s)}=
    \pm ie\varepsilon \Psi_\pm^{\mu(s)}\,,\qquad  \delta_\varepsilon \Lambda_\pm^{\mu(s-1)}=\pm ie\varepsilon \Lambda_\pm^{\mu(s-1)}\,,
    \end{equation}
we find $r_1=8s-3$. For Lagrangian equations of motion, gauge identities are given by the Noether identities. The latter are generated by the formal transpose of the gauge-symmetry generators (\ref{R1}) and (\ref{R2}). Since all the equations of motion are first order, we readily conclude that $t_1=24s+11$ and $l_2=8s-3$. By formula (\ref{N}), 
\begin{equation}
    \mathcal{N}=t_1-r_1-2l_2=20\,.
\end{equation}
Of the 10 real polarizations, two belong to the subsystem of massless fields $B,A,\varphi$, and the remaining $8$ are equally distributed between the  fields of the original and dual systems. 

In the above calculation, we considered the vector field $B$ to be real, which is not entirely accurate due to the complex sources on the r.h.s. of Eqs. (\ref{DS}). By complexifying the vector field $B$, we increase the number of (real) polarizations by one.  The real-valued $B_\mu$, however, is admissible at the free level. For $e=0$, the dual system simplifies to:
\begin{equation}\label{FDS}
 (W\mp sm)\Psi_\pm^{\mu(s)}-\partial^{\mu} \Lambda_\pm^{\mu(s)} = 0\,,\qquad  WB_\mu=0\,, \qquad \partial_\mu B^\mu=0\,.    
\end{equation}
The gauge invariance
\begin{equation}
    \delta_\xi\Psi^{\mu(s)}_\pm=\partial^{\mu}\xi^{\mu(s-1)}_\pm \,,\qquad \delta_\xi \Lambda^{\mu(s-1)}_\pm=(W\mp sm)\xi^{\mu(s-1)}_\pm
\end{equation}
allows us to impose the following gauge-fixing conditions:
\begin{equation}\label{G}
    \partial_{\nu}\Psi^{\nu\mu(s-1)}_\pm=0\,,\qquad \Lambda^{\mu(s-1)}_\pm=0\,.
\end{equation}
In this gauge, the first equation in (\ref{FDS}) describes a massive particle of helicity $\pm s$. As for the remaining two equations, they resemble the equations for the Chern--Simons field $B$ in the Lorentz gauge. The equation $WB_\mu=0$ enjoys an obvious gauge identity, and applying formula (\ref{N}) gives one propagating polarization, as expected.  

We see that the complex Lagrangian (\ref{L}) describes a consistent propagation of two massive spin-$s$ particles coupled to the Bogomolny background.  In the gauge (\ref{G}), each particle is described by the doublet of fields $(\Phi^\pm_{\mu(s)}, \Psi_{\mp}^{\mu(s)})$ with the same electric charge and opposite helicities. 

Notice that the gauge transformations (\ref{R1}) do not affect the fields $\Phi_\pm$. Therefore, we can augment the Lagrangian with the $U(1)$-invariant term $\Phi^{+\mu(s)}\Phi^-_{\mu(s)}$ without changing the total number of physical degrees of freedom.  Excluding the fields $\Phi^\pm_{\mu(s)}$
from (\ref{L}) by their equations of motion, we obtain a second-order Lagrangian for the remaining fields. Explicitly, 
\begin{equation}\label{L2}
\begin{array}{rcl}
    L_2&=&\big[ (W_A-sm-ise\varphi)\Psi_+ ^{\mu(s)}-D^{\mu}\Lambda_+^{\mu(s-1)}\big] \\[3mm]
    &\times &\big[ (W_A+sm+ise\varphi)\Psi_-{}_{\mu(s)}-D_{\mu}\Lambda_-{}_{\mu(s-1)}\big]\\[3mm]
    &+&B^\mu (WA_\mu -\partial_\mu \varphi)\,.
    \end{array}
    \end{equation}
The Lagrangian is still invariant under the gauge transformations (\ref{R1}) and (\ref{R2}), where the gauge variation of $B^\mu$ is now given by
\begin{equation}
    \delta_\xi B_{\mu} = ies\xi_{\pm}^{\mu(s-1)}\big[\big(W_A\pm s(m + ie\varphi)\big)\Psi^{\mp}_{\mu(s)} - D_{\mu}\Lambda^{\mp}_{\mu(s-1)}\big]\,.
\end{equation}
The equations of motion for matter fields take the form
\begin{equation}\label{2Int}
\begin{array}{c}
\displaystyle   D^2\Psi_\pm^{\mu(s)}+ M^2\Psi_\pm^{\mu(s)} \mp \frac{ie}{s} F^{\lambda \mu}\Psi_\pm{}_\lambda{}^{\mu(s-1)}=\frac1{s^2}(W_A\pm sm\pm ise\varphi)D^{\mu}\Lambda_\pm^{\mu(s-1)}\,, \\[5mm]
   D_{\mu}\big[ (W_A\mp sm\mp ise\varphi)\Psi_\pm ^{\mu(s)}-D^{\mu}\Lambda_\pm^{\mu(s-1)}\big]=0\,,
  \end{array}
\end{equation}
where  $ D_\mu=\partial_\mu + ieA_\mu$ and $M=m+ie\varphi$.  To fully elucidate the physical spectrum of the model, we can turn off the interaction for a while by setting $e=0$. Then, in the admissible gauge 
\begin{equation}
    \Lambda_\pm^{\mu(s-1)}=0\,,\qquad \partial_{\nu}\Psi^{\nu\mu(s-1)}_\pm=0\,,
\end{equation}
the free equations of motion for the matter fields reduce to the mass-shell condition $(\square +m^2)\Psi^{\mu(s)}_\pm=0$. Hence, each of the two fields $\Psi^{\mu(s)}_\pm$ describes a particle of mass $m$ and spin $s$. 
Returning to the interacting system, we can impose the partial gauge $\Lambda^{\mu(s-1)}_\pm=0$, fixing the auxiliary fields $\Lambda_\pm^{\mu(s-1)}$. 
Then, the first equation in (\ref{2Int}) tells us that the massive particles have opposite electric charges and a gyromagnetic ratio of $1/s$.

It is important to emphasize that although Eqs. (\ref{2Int}) were derived from the complex Lagrangian (\ref{L2}), they are fully consistent as long as the background fields $A_\mu$ and $\varphi$ satisfy the Bogomolny equations (\ref{BE}). Formally, these second-order equations are obtained from the chiral system (\ref{Int2}) by the substitution $\Phi^\pm{}^{\mu(s)}=(W_A\pm sm\pm ise\varphi)\Psi_\mp ^{\mu(s)}-D^{\mu}\Lambda_\mp^{\mu(s-1)}$.
This replacement, while preserving the gauge identities, introduces an additional gauge symmetry, so that the total number of physical degrees of freedom doubles.

\section{Dimensional reduction of higher-spin SDYM}\label{S5}
In this section, we reveal the secret to deriving all of the above equations in one fell swoop. Specifically, we will show that each of them can be obtained through the dimensional reduction\footnote{For the first time the idea of dimensionally reducing chiral higher-spin gravity to obtain a theory with massive higher-spin fields in $3d$ was explored in the light-cone gauge in \cite{Metsaev:2020gmb,Skvortsov:2020pnk}. } of a higher-spin extension of the self-dual Yang-Mills theory \cite{Krasnov:2021nsq}, which itself is a truncation \cite{Ponomarev:2017nrr} of chiral higher-spin gravity \cite{Metsaev:1991mt, Metsaev:1991nb, Ponomarev:2016lrm, Skvortsov:2018jea, Skvortsov:2020wtf} in $4d$ Euclidean space.

We begin by examining  the cases of spin $0$ and $1$, where the relevant chiral theory is described by the action:
\begin{equation}\label{CS}
    S=\int_{\mathbb{R}^4} B\wedge F\,.
\end{equation}
As above, $F=dA$ is the strength tensor of the gauge vector field $A$ and $B$ is an anti-self-dual two-form. This theory can be viewed as an Abelian version of the self-dual Yang--Mills (SDYM) theory introduced by Chalmers and Siegel in \cite{Chalmers:1996rq}. 
The action is invariant under standard gauge transformations, $A\rightarrow A+ d\lambda$, and yields the following equations of motion: 
\begin{equation}\label{FB}
 F=\ast F\,,   \qquad  dB=0\,.
 \end{equation}
The field $B$ acts as a Lagrange multiplier, imposing the self-duality condition on the strength tensor $F$.  By adding a gauge-invariant term $B\wedge B$ to the action (\ref{CS}) and integrating out $B$, we recover the Maxwell action.  Consequently, the equations of motion (\ref{FB}) describe two propagating polarizations, which correspond to the two photon states. As both polarizations of a massless particle are present, the terminology ``chiral'' and ``self-dual'' may appear somewhat misleading in this context. These terms are used to emphasize the asymmetric treatment of the right- and left-handed polarizations: one is represented through the gauge potential $A$, while the other is described by the field strength $B$.

Let us now compactify one of the four dimensions, specifically $x_4$, on a circle of circumference $2\pi r$. Each component of the one-form field $A$ and the two-form field $B$ can then be expanded into a Fourier series as follows:
\begin{equation}\label{FE}
   A=\sum_{n\in \mathbb{Z}}A^{(n)}e^{\frac{i n x_4}{r}} \,, \qquad   B=\sum_{n\in \mathbb{Z}}B^{(n)}e^{\frac{ i n x_4}{r}} \,.
   \end{equation}
In this expression, the complex amplitudes $A^{(n)}$ and $ B^{(n)}$ depend only on the remaining coordinates $x^\mu$, where $\mu=1,2,3$. 
The first equation in (\ref{FB}) reduces to the following form:
\begin{equation}
    F_{\mu 4}=\frac12\varepsilon_{\mu\nu\lambda}F_{\nu\lambda}\,.
\end{equation}
By substituting the Fourier expansion (\ref{FE}) for $A$, we derive an infinite set of 3d equations for the complex amplitudes, specifically:
\begin{equation}\label{FF}
    W A^{(n)}_\mu= \partial_\mu \varphi^{(n)} -   M_n A^{(n)}_\mu\,,
\end{equation}
where $\varphi^{(n)}=A_4^{(n)}$  and $ M_n={ in}/{r}$. This system exhibits an evident gauge symmetry, give by:
\begin{equation}
A^{(n)}_\mu \rightarrow A^{(n)}_\mu+ \partial_\mu \lambda^{(n)}\,,\qquad\varphi^{(n)}\rightarrow \varphi^{(n)}+M_n\lambda^{(n)}\,.
\end{equation}
For $n=0$, $M_0=0$ and we recover the Bogomolny equation (\ref{BE}) with a single propagating polarization. 
The fields $\varphi^{(n)}$ with $n\neq 0$ resemble the Stueckelberg fields: they can be eliminated through the field redefinition $A_\mu^{(n)}\rightarrow A_\mu^{(n)}-M_n^{-1}\partial_\mu\varphi^{(n)}$. Then Eqs. (\ref{FF}) take the form of massive Chern--Simons theory
\begin{equation}
    WA^{(n)}_\mu=  -  M_n A^{(n)}_\mu\,,\qquad n\neq 0\,.
\end{equation}
This system is not involutive and has a nontrivial differential consequence: $\partial^\mu A_\mu^{(n)}=0$. Using the general formula for counting physical degrees of freedom, we find that threre is one (complex) propagating polarisation.  

The $3+1$ decomposition of the two-form field $B$ gives $B=(B_{\mu 4}, B_{\mu\nu})$. Anti-self-duality implies that $B_{\mu 4}=B_\mu$, where $B_\mu=-\frac12\epsilon_{\mu\nu\lambda}B_{\nu\lambda}$. The second equation in (\ref{FB}) leads to the following  system for the Fourier components:
\begin{equation}
    \partial^\mu B^{(n)}_\mu=0\,,\qquad W B^{(n)}_\mu=M_n B^{(n)}_\mu\,.
\end{equation}
For $n\neq 0$, we recover the equations of massive Chern--Simons theory as above. In the case $n=0$, we obtain massless Chern--Simons equations in the Lorentz gauge.

Integrating over the circle in (\ref{CS}) yields the action for the Fourier modes:
\begin{equation}
    S=8\pi r\int_{\mathbb{R}^3}\Big[ B^{(0)}{}^\mu (WA^{(0)}_\mu -\partial_\mu\varphi^{(0)})+\sum_{n\in \mathbb{Z}\backslash \{0\}} B^{(-n)}{}^\mu (WA^{(n)}_\mu +M_n A^{(n)}_\mu -\partial_\mu\varphi^{(n)}) \Big]d^3x\,.
    \end{equation}
Thus, we see that the zero modes of the fields $A$, $\varphi$ and $B$ exactly reproduce the dynamics of massless fields in (\ref{L}), while the nonzero modes are identified with the free massive fields of helicity $\pm1$: 
\begin{equation}
\begin{array}{cccc}
   B^{(-n)\mu} =\Phi_+^{\mu} \,,&\qquad  A_\mu^{(n)}=\Psi^+_\mu\,,&\qquad  \varphi^{(n)}=\Lambda_+\,,&\qquad n>0\,;\\[3mm]
    B^{(-n)\mu} =\Phi_-^{\mu} \,,&\qquad  A_\mu^{(n)}=\Psi^-_\mu\,,&\qquad  \varphi^{(n)}=\Lambda_-\,,&\qquad n<0\,.  
    \end{array}
\end{equation}
Note that the resulting mass spectrum, $M_n=in/r$, is purely imaginary.  This is a consequence of the Euclidean signature of the metric. To make contact with the formulas from the previous section, we can perform the Wick rotation:
\begin{equation}
    x^1\longrightarrow ix^1\,, \qquad A_1^{(n)}\longrightarrow i A_1^{(n)}\,.
\end{equation}
This effectively replaces $M_n$ with $ n/r$ and transforms the Euclidean metric to the Lorentz metric.

The Chalmers--Siegel action (\ref{CS}) allows for a higher-spin extension, as shown in \cite{FLORES2017431}, \cite{krasnov2021actions}. This extension is most naturally expressed in terms of the two-component spinor formalism. Recall that all finite-dimensional irreducible representations of the orthogonal group $SO(4)=SO(3)\times SO(3)$ are parametrized by a pair $(n,m)$ of natural numbers and can be realized in the space of spin-tensors 
\begin{equation}\label{AA'}
    \psi^{A_1\cdots A_n, A'_1\cdots A'_m}
\end{equation}
which are fully symmetric in the primed and unprimed groups of indices. All indices take only two values, $0$ and $1$,  and transform under the spinor representation of the rotation group $SO(3)$.  They can be raised and lowered with the help of the $SO(3)$-invariant spin-tensors $\epsilon_{AB}=-\epsilon_{BA}$ and $\epsilon^{AB}=-\epsilon^{BA}$. By definition, $\xi^A=\epsilon^{AB}\xi_B$ and $\xi_A=\xi^B\epsilon_{BA}$ with the same relations holding for the primed indices.  Each $SO(4)$-vector $v^i=\delta^{ij}v_j$ corresponds to a bi-spinor $v_{AA'}=\sigma_{AA'}^i v_i$ and vice versa; here, $\sigma_{AA'}^i$ are the van der Waerden symbols that satisfy the identity $\delta_{ij}\sigma^i_{AA'}\sigma^j_{BB'}=\epsilon_{AB}\epsilon_{A'B'}$. 
As usual, we use the shorthand notation $\psi^{A(n),A'(m)}$ for the spin-tensors (\ref{AA'}). The advantage of the spinor formalism over the vector formalism is that it makes all (anti-)self-duality conditions explicit. 

The higher-spin extension of the Chalmers--Siegel action now reads \cite{Krasnov:2021nsq}
\begin{align}\label{HS-CS}
   S= \int \Psi^{A(2s)} \pl\fdu{A}{A'} \Phi_{A(2s-1),A'} \;d^4x\,.
\end{align}
The action is invariant under the gauge symmetry transformations 
\begin{align}\label{gtr}
    \delta_\xi \Phi_{A(2s-1),A'}&= \pl_{AA'} \xi_{A(2s-2)}\,.
\end{align}
and so are the corresponding equations of motion
\begin{equation}\label{EoM}
     \partial^{B}{}_{A'}\Psi_{A(2s-1)B}=0\,,\qquad \partial_A{}^{A'}\Phi_{A(2s-1),A'}=0\,.
\end{equation}
The spin-tensor fields $\Psi^{A(s)}$ and $\Phi_{A(s-1),A'}$ describe the states of a massless particle with helicities  $+s$ and $-s$, respectively.   
By construction, the action (\ref{HSCS}) is complex. However, as noted in \cite{FLORES2017431}, 
its real part leads to the same equations of motion. For example, setting $s=1$
 and taking the real part, we arrive at the Chalmers--Siegel action (\ref{CS}). The vector-spinor dictionary implies the following 
 identification of fields:
 \begin{equation}
     B\longleftrightarrow \Psi_{AA}\,, \qquad A\longleftrightarrow \Phi_{AA'}\,.
 \end{equation}

To perform dimensional reduction, we compactify the fourth coordinate  on a circle of radius $r$, thereby reducing the 
$SO(4)$ symmetry to its diagonal subgroup $SO(3)\subset SO(3)\times SO(3)$. 
Upon restriction to $SO(3)$, all chiral representations $(m,0)$ and $(0,n)$ remain irreducible, while mixed-type representations are decomposed into multiple irreducible parts. In particular
\begin{align}
    \Phi_{A(2s-1),A'}&= \Phi_{A(2s-1)A'}+\epsilon_{AA'}\Lambda_{A(2s-2)}=\big(\Phi_{A(2s)}, \Lambda_{A(2s-2)}\big)
\end{align}
and 
\begin{align}
    \pl_{AA'}= d_{(AA')} +i\epsilon_{AA'} \pl_4\,, \qquad d_{AA}=\sigma^\mu_{AA}\partial_\mu\,,\qquad \mu=1,2,3\,,
\end{align}
where $\sigma^\mu_{AA}$ are the Pauli matrices for $SO(3)$. The action (\ref{HS-CS}) now takes the form 
\begin{align}\label{HSCS}
   S = \int \Big[ d^{BB}\Psi_{A(2s - 2)BB}\Lambda^{A(2s - 2)}+i\pl_4\Psi^{A(2s)}\Phi_{A(2s)}-\Psi^{A(2s-1)B}d\fdu{B}{C}\Phi_{A(2s-1)C}\Big] d^4x\,.
\end{align}
Integrating over the circle yields the action for the Fourier modes of the fields, producing an infinite mass spectrum $M_n=in/sr$ for $n=0,1,2,\ldots$, associated with the eigenvalues of the operator $i\partial_4$. The corresponding equations of motion are 
\begin{equation}\label{E1}
\begin{array}{c}
 d_{A}{}^{B} \Phi_{A(2s-1)B}+ i\pl_4 \Phi_{A(2s)}-d_{AA}\Lambda_{A(2s-2)}=0\,,\\[3mm] 
 d^{B}{}_{A}\Psi_{A(2s-1)B}+i\pl_4 \Psi_{A(2s)}=0\,,\qquad
    d^{BB}\Psi_{A(2s-2)BB}=0\,.
    \end{array}
\end{equation}
They are invariant under the following gauge transformations:
\begin{align}
    \delta_\xi \Phi_{A(2s)}= d_{AA} \xi_{A(2s-2)}\,,\qquad 
    \delta_\xi \Lambda_{A(2s-2)}= d_A{}^{B} \xi_{A(2s-3)B}+i\pl_4 \xi_{A(2s-2)}\,.
\end{align}
Again, making the identifications 
\begin{equation}
    \Psi_{\mu(s)}\longleftrightarrow \Psi_{A(2s)}\,,\qquad  \Phi_{\mu(s)}\longleftrightarrow \Phi_{A(2s)}\,,\qquad \Lambda_{\mu(s-1)}\longleftrightarrow \Lambda_{A(2s-2)}\,,\qquad    \end{equation}
and passing to the Fourier modes, one can recognize in system (\ref{E1}) various equations (\ref{dF}, \ref{Wm}) and (\ref{FDS}) for massive and massless fields in three dimensions. 

The free actions (\ref{HSCS}) allow us to introduce a minimal interaction between massless particles of spin $1$ and $s$. This is achieved by interpreting the fields $\Phi_{A(2s)}$ and $\Psi_{A(2s)}$ as sections of $U(1)$ bundles over $\mathbb{R}^4$ that are dual to each other. With this interpretation, the gauge vector field $\Phi_{AA'}$ is identified as a $U(1)$-connection. The minimal coupling is then introduced  by modifying the free action (\ref{HSCS}) to
\begin{align}
    \label{Act}
    S=\int \Big[\Psi^{A(2s)} D\fdu{A}{A'} \Phi_{A(2s-1),A'} + \Psi^{AA}\partial_A{}^{A'}\Phi_{AA'}\Big]d^4x \,,
\end{align}
where $D_A{}^{A'}=\partial_A{}^{A'}+ie \Phi_A{}^{A'}$ represents the $U(1)$-covariant derivative, and the second term corresponds to the Chalmers--Siegel action (\ref{CS}) expressed in spinor form.
By construction, the action enjoys the standard $U(1)$ local symmetry: 
\begin{equation}
\delta_\varepsilon \Phi_{AA'}=\partial_{AA'}\varepsilon\,,\qquad \delta_{\varepsilon}\Phi_{A(2s-1),A'}=-ie\varepsilon \Phi_{A(2s-1),A'}\,,\qquad \delta_{\varepsilon}\Psi^{A(2s)} = ie\varepsilon\Psi^{A(2s)}\,.
\end{equation}
It also remains invariant under the gauge transformations for spin-$s$ fields:
\begin{equation}
    \delta_\xi \Phi_{A(2s-1),A'}=D_{AA'}\xi_{A(2s-2)}\,,\qquad \delta_\xi \Psi^{AA}=-\Psi^{A(2s)}\xi_{A(2s-2)}\,.
\end{equation}
It is now evident that, upon dimensional reduction, this action reproduces a sum of actions for 3d massive fields coupled to the Bogomolny background associated with the zero modes of $\Phi_{AA'}$. 

The action (\ref{Act}) admits a straightforward non-Abelian generalization. Instead of the gauge group $U(1)$, one can consider an arbitrary Lie group $G$, with the fields $\Psi^{A(2s)}$ and $\Phi_{A(2s-1),A'}$ taking values in dual representations of
$G$. With this modification, the first term in (\ref{Act}) retains its form, while the second term is replaced by
\begin{equation}
    \Psi^{AA}\big(\partial_A{}^{A'}\Phi_{AA'} +ie[\Phi_A{}^{A'}, \Phi_{AA'}]\big)\,.
\end{equation}
Here, square brackets denote the commutator in the Lie algebra of $G$. 

Another non-Abelian generalization of the free action was recently proposed in \cite{adamo2023higher}. In that paper, the authors consider an infinite set of massless fields $\{\Psi^{A(2s)}, \Phi_{A(2s-1),A'}\}_{s=1}^\infty$ taking values in a semi-simple Lie algebra. They introduce the higher-spin counterpart of the SDYM curvature,
\begin{equation}
    F_{A(2s)}=\partial_A{}^{A'}\Phi_{A(2s-1),A'}+\sum_{r+t=s+1}\big [\Phi_{A(2r-1)}{}^{A'}, \Phi_{A(2t-1),A'}\big ]\,,
\end{equation}
and define the action by the integral
\begin{equation}\label{SDYM}
    S[\Psi, \Phi]=\sum_{s=1}^\infty \int_{\mathbb{R}^4}d^4x\,\mathrm{tr}\big( \Psi^{A(2s)}F_{A(2s)}\big)+\frac{\mathrm{g}^2}4\int_{\mathbb{R}^4}d^4x\,\mathrm{tr}
    \big (\Psi^{A(2s)}\Psi_{A(2s)}\big)\,.
    \end{equation}
This action admits various truncations. First, we can set $\mathrm{g}=0$ or integrate out  the $\Psi$'s from the action. Second, the sum (\ref{SDYM}) can be truncated at any given spin $s$, retaining only a finite number of fields. Moreover, it is possible to keep only the spin-$1$ and spin-$s$ fields. 
For $\mathrm{g}=0$, this latter option reproduces the action (\ref{Act}) for $\Psi^{A(2s)}$ and $\Phi_{A(2s-1),A'}$ in the adjoint representation.  

By reducing the non-Abelian actions above to three dimensions, one obtains non-linear equations of motion that describe a consistent interaction of massive and massless higher-spin fields.   However, a detailed  exploration of the general interaction problem for higher-spin fields in $d=3$ is beyond the scope of this paper.

\section{Discussion and conclusions} \label{DC}
Although there are no fundamental obstructions to introducing various interactions of massive higher-spin fields, at least classically,  there is currently no efficient formalism available to do this\footnote{At the quantum level, higher-spin interactions are expected to be non-renormalizable when taken individually, a possible solution being to introduce an infinite tower of fields of ever-growing spin as in string theory.}. Starting from the standard description based on symmetric tensors, the main challenge is to exclude longitudinal states. This encourages the search for new descriptions of massive fields with spin, such as the recently proposed chiral formulation \cite{Ochirov:2022nqz} and its generalization \cite{Delplanque:2024bmn}, as well as new ways to decouple the longitudinal modes, including the Stuckelberg (massive) gauge symmetry \cite{Zinoviev:2001dt}, which is also a key component of the string-inspired BRST description \cite{Pashnev:1997rm}.  

In this paper, we proposed new actions for massive fields with spin. Additionally, we explored the potential to introduce electromagnetic interactions with the Bogomolny background, which can also be obtained by the dimensional reduction of the higher-spin extension of self-dual Yang--Mills theory \cite{Ponomarev:2017nrr,Krasnov:2021nsq} which is itself a truncation of chiral higher-spin gravity \cite{Metsaev:1991mt, Metsaev:1991nb, Ponomarev:2016lrm, Skvortsov:2018jea, Skvortsov:2020wtf}. Self-dual theories exhibit many useful features, including integrability and UV-finiteness (see, e.g. \cite{Krasnov:2016emc} for a discussion of the latter). Moreover, all amplitudes and solutions of self-dual theories are part of the complete theories for which they are self-dual truncations.

We do not expect chiral higher-spin gravity (or its truncations) to have a perturbatively local completion \cite{Ponomarev:2017nrr}, but this argument does not apply to massive fields in three dimensions. On the contrary, one would expect most of the interactions that are consistent at low orders to have some higher-order completion. One advantage of the interactions introduced in this paper is that they are complete, meaning no higher-order corrections are needed. By contrast, within the symmetric tensor approach, interactions that are consistent at the cubic order require higher-order corrections, and the expansion continues indefinitely.

One drawback of any chiral approach (see, e.g. \cite{Ochirov:2022nqz, Delplanque:2024bmn}) is that discrete symmetries, such as the parity symmetry, are not straightforward to incorporate. One has to impose parity order by order, which, nevertheless, seems to be quite efficient \cite{Cangemi:2023bpe}. It would be important to understand how discrete symmetries can be imposed systematically in chiral approaches. The actions for massive fields proposed in this paper are very reminiscent of the chiral ones and the $3d$ seems to be a useful playground. 

In summary, this work addresses the challenges inherent in formulating interactions for massive higher-spin fields, emphasizing the limitations of existing approaches while proposing new avenues for progress. By introducing novel actions and equations for massive higher-spin fields in $3d$ and exploring their connections to electromagnetic interactions and self-dual Yang--Mills theory, we have demonstrated the potential for constructing  all-order consistent interactions in three dimensions. While chiral formulations provide an elegant framework, challenges such as the incorporation of discrete symmetries, remain an open question. Future research will aim to refine these formulations and explore their broader implications, particularly in the context of three-dimensional theories coupled to a gravitational background. 
\section*{Acknowledgements} 

The work of E.S. was partially supported by the European Research Council (ERC) under the European Union’s Horizon 2020 research and innovation programme (grant agreement No 101002551).

\end{document}